# Elementary analysis of galaxy clusters:
# similarity criteria, Tully–Fisher, and approximate invariants

## Georgiy S. Golitsyn

### A.M.Obukhov Institute of Atmospheric Physics, RAS, Moscow 117019

### email: gsg@ifaran.ru

## Abstract

At observations of galaxy clusters the following quantities are usually measured: luminosity $L$, size $R$, mass $M$, temperature $T_e$, sometimes velocities. They all are interbinded by the gravity field characterized by the universal constant G. These five quantities are determined by three measurements units: mass M, length L and time T. Therefore one can form two non-dimensional similarity criteria: $\Pi_1$ and $\Pi_2$. One can also form any chosen observable as a function of the other three ones. The author has at hand the data by Vikhlinin (2002) and Vikhlinin et al. (2006), rather more complete than any other. This material consists of more than thirty clusters at $0.4 \leq z < 1.26$ and $z \leq 0.23$ and represents various stages of relaxation. This material gives a possibility to test the derived the dimensional relationships and to determine the dimensionless numerical coefficients at these relationships. These coefficients are found nearly constant with a scatter less than 30 per cent in the data above and could be considered as other similarity criteria but functions of $\Pi_1$ and $\Pi_2$. With such a small scatter they may be called approximate invariants. The luminosity $L$ and universal constant $G$ are forming the dynamical velocity scale $U_d$, which immediately explains the empirical Tully – Fisher law as: $L \approx U^5/G$. Having the temperature $T_e$ one may determine the thermal velocity of the gas plasma particles $U_T$. The ratio $U_d/U_T = \Pi_1$ is used here as a new similarity criterium which is found to be constant within six per cent for nearly 30 objects cited above: $\Pi_1 = 0.163 \pm 0.009 \approx 1/6$ and may be interpreted as the Mach number. The other criterium $\Pi_2$ is the ratio of the cluster potential energy to its doubled kinetic one and is the virial one. It is found to be a function of the cluster age. This is an evidence of cluster evolution during their life time, evidently through "cannibalism" of neighbours. At $z > 0.5$ the mean cluster mass is five times less, that at small $z \leq 0.2$. It is demanding to expand these results to other clusters and different objects: singular galaxies, stars and their clouds, etc. As an example it is found that for the Sun $\Pi_1 = 0.078$, only about a half of the cluster values despite 14-15 orders of magnitude difference in mass and 4-5 magnitude difference in radiation temperature.

## 1. Introduction

The galactic clusters consist of many dozens, hundreds and even thousands individual galaxies. It is one of the fastest developing areas in astrophysics, both in theory and observations, surface and space ones. Their formation and characteristics are related to the processes of the Universe formation, its structure, geometry, and time evolution. The two reviews, Voit (2005) and Kravtsov and Borgani (2012), each with about 400 references, present a good introduction into the subject and the state of knowledge in this direction. There are also analytical models, e.g. for temperature and density radial profiles of intergalactic gas which have a self-similar character, i.e. the similar behavior for different clusters. This is confirmed by observations of clusters with small red shift $z \leq 0.2$ and by sizes near 1 Mpc= $3.086 \cdot 10^{22}$ m. Such clusters are observable with a distinct spatial resolution. The data of measurements reveal a numbers of relations among their characteristics such as mass – luminosity, mass – temperature, luminosity – temperature, luminosity – velocity for clusters and separate galaxies (e.g. Tully–Fisher, 1977), etc. The numerical models using billions of sample particles bound gravitationally, describing star formation, cooling by radiation and a number of other effects can reproduce some of observationally found relationships, but far from all ones.

E.g., the hydrostatic approximations in the centrally – symmetric gravitationally field produces that mass of the gas objects in the proportional to its temperature $T_e^{3/2}$ for the temperatures in the range approximately from 2 to 10 keV, 1 keV = $1.161 \cdot 10^7$ K, i.e. from 20 to 100 millions K. At the smaller temperatures the exponent can be larger than 1.5, and therefore in dependence on the temperature range used the exponent may be found larger, see e.g. Saunderson et al (2003), Vikhlinin et al (2006).

At the same time, the measurable cluster characteristics are strictly determined physical values characterized be corresponding dimension. All relationships between them must maintain the dimensional rules. The dimensions of our observable are: luminosity $[L] = ML^2T^{-3}$, size $[R] = L$, mass $[M] = M$, velocity $[U] = LT^{-1}$, $G = 6,672 \cdot 10^{-12}$ m³kg⁻¹s⁻², $[G] = L^3M^{-1}T^{-2}$. Altogether we have 5 quantities determined by 3 measurements units. According to celebrated Π-theorem by Edgar Buckingham (1914) in such a situation there are two (5-3=2) non-dimensional combinations among our five quantities. Their determination will be performed below and now we shall invoke two scales from the book by Martin Rees (2002):

$$T_d = (\rho G)^{-1/2}, \tag{1}$$

where ρ is the gas density. The second scale of thermal velocity is introduced by mean temperature $T_e$ equating it to the particle kinetic energy as:

$$3k\, T_e = \mu m_\rho U_T^2, \tag{2}$$

where $U_T$ is the particle thermal velocity, $\mu = 0.6$ is the mean molecular weight of the gas particle, consisting of electrons, protons and of ten per cent of helium, $k=1.381\cdot10^{-23}$ J/K is the Boltzmann's constant. $T_e$ here is in Kelvin degrees and $1\text{keV} = 1.161\cdot10^7$K.

If the cluster mass is known, and mass fraction is of about 10 per cent (Kravtsov and Borgani, 2012), then one can estimate the number of the gas particle as:

$$N \leq 0.1 \; M/\mu m_p, \tag{3}$$

And then the enthalpy, or the gas heat content is

$$E_T = 0.3 \; k \; T_e M(\mu m_p)^{-1} \tag{4}$$

Vikhlinin et al (2002, 2006) for clusters use a very successful and mutual concordant system of the measurements units: for mass $10^{14}$ M$_\odot$, with the solar mass M$_\odot$=$2\cdot10^{30}$ kg; for luminosity $10^{37}$ W (the solar luminosity is $4\cdot10^{26}$ Watt); the size is in megaparsec, 1 Mpc = $3.086\cdot10^{22}$m. This leads to the values of the measured parameters of order unity, as well as for non-dimensional coefficients arising in the dimensional equations. This will be seen below. The basis for the dimensional analysis can be found at Bridgman (1931), Sedov (1959), Barenblatt (1996, 2003).

## 2. Dimensional analysis

For cluster we have a number of measured characteristics which allows us to introduce several scales. The time scale (1) can be written as:

$$T_d = R^{3/2}(MG)^{-1/2}. \tag{1'}$$

With the luminosity we may introduce the velocity scale:

$$U_d = (LG)^{1/5} \tag{5}$$

And we shall call it the dynamical velocity scale. Measuring the velocity in $10^{37}$ W we obtain, remembering the universality of the constant $G$ that

$$U_d = 146 \; L^{1/5} \tag{5'}$$

where the velocity occurs to be in km/s, i.e. this scale is in hundreds kilometers per second. In the same measuring units the thermal velocity will be as:

$$U_T = 693 \; T_e^{1/2}, \; \text{km/s}. \tag{6}$$

The eqs. (2) - (6) introduce the scale of energy, more precisely, of enthalpy which in our units is equal to:

$$E_0 = 0.96\cdot10^{55}M \; T_e \approx 1\cdot10^{55} \; MT_e, \; \text{Jouls}. \tag{4'}$$

The presence of the two velocity scale allows one to propose a new similarly criterium:

$$\Pi_1 = U_d/U_T \tag{7}$$

which may be called the Mach number, because the sound velocity in gas is close, a little bit less, than the gas particle thermal velocity. The eq. (5) immediately explains the Tully–Fisher (1977) relation: $L \sim U^5G^{-1}$, known for nearly forty years, but still waiting for an explanation (and a theory based on "the first principles").

The second similarity criterium is well known for astronomers as one of the fundamental results of analytical mechanics (Landau and Lifshits, 1960): the virial relationship. For the centrally symmetric situation it can be written as

$$\Pi_2 = MGU^{-2}R^{-1}, \tag{8}$$

And this quantity must be negative and equal to -1, because the system is in a potential well. The eq. (8) is the ratio of the potential (negative) energy to the doubled kinetic energy. The total energy of the system, i.e. the sum of the kinetic and potential energy must be negative, otherwise some parts of the system, i.e. of a cluster, will go to infinity. The virial relationship is established in the system asymptotically after a long time (Landau and Lifshits, 1960). To estimate such a time it should be measured in dynamical time periods $T_e$, eq. (1).

The concrete values of the cluster parameters are in works by Vikhlinin (2002), by Vikhlinin et al (2006, 2010). We present two Tables from those works of 2002 and 2006 which are edited and added by this author.

Table 1. Parameters of the distant clusters of 2002,

the similarity criteria $\Pi_1$ and $\Pi_2$ and characteristics of their ages.

| № | $z$ | $T_e$, keV | $L$, $10^{37}$W | $M$, $10^{14}M_\odot$ | $R$, Mpc | $\Pi_1$ | $\Pi_2$ | $T_a$ By | $T_d$ By | $T_a/T_d$ |
|---|---|---|---|---|---|---|---|---|---|---|
| 1 | 0.394 | 4.8 | 9.2 | 1.24 | 0.5 | 0.150 | 0.96 | 8.12 | 0.48 | 16.9 |
| 2 | 0.400 | 3.7 | 8.9 | 1.42 | 0.7 | 0.170 | 1.02 | 8.09 | 0.49 | 16.5 |
| 3 | 0.424 | 3.6 | 10.6 | 1.07 | 0.5 | 0.178 | 1.05 | 7.92 | 0.61 | 13.0 |
| 4 | 0.426 | 7.6 | 27.0 | 2.89 | 0.9 | 0.148 | 0.79 | 7.91 | 0.50 | 15.8 |
| 5 | 0.451 | 14.1 | 260.4 | 8.77 | 0.9 | 0.171 | 1.23 | 7.75 | 0.43 | 18.0 |
| 6 | 0.453 | 5.8 | 15.9 | 1.81 | 0.7 | 0.152 | 0.83 | 7.73 | 0.65 | 11.9 |
| 7 | 0.460 | 5.3 | 16.3 | 1.57 | 0.5 | 0.160 | 1.10 | 7.68 | 0.55 | 14.0 |
| 8 | 0.516 | 5.1 | 15.7 | 1.67 | 0.6 | 0.162 | 1.02 | 7.34 | 0.54 | 13.6 |
| 9 | 0.537 | 8.1 | 91.7 | 3.68 | 1.0 | 0.183 | 0.85 | 7.21 | 0.78 | 9.2 |
| 10 | 0.541 | 9.9 | 113.3 | 6.43 | 1.0 | 0.173 | 1.21 | 7.19 | 0.59 | 12.2 |
| 11 | 0.562 | 4.8 | 12.5 | 1.19 | 0.5 | 0.160 | 0.92 | 7.07 | 0.49 | 14.4 |
| 12 | 0.574 | 2.7 | 38.8 | 0.36 | 0.5 | 0.167 | 0.50 | 7.00 | 0.88 | 8.0 |
| 13 | 0.583 | 5.2 | 10.8 | 0.95 | 0.5 | 0.149 | 0.68 | 6.95 | 0.54 | 12.9 |
| 14 | 0.700 | 7.2 | 28.7 | 2.01 | 0.7 | 0.154 | 0.74 | 6.36 | 0.62 | 10.2 |
| 15 | 0.782 | 6.3 | 32.4 | 1.41 | 0.7 | 0.168 | 0.59 | 5.99 | 0.74 | 8.1 |
| 16 | 0.805 | 2.2 | 2.0 | 0.21 | 0.5 | 0.163 | 0.36 | 5.89 | 1.16 | 5.1 |
| 17 | 0.805 | 4.3 | 13.2 | 1.04 | 0.8 | 0.170 | 0.56 | 5.89 | 1.05 | 5.6 |
| 18 | 0.813 | 6.6 | 28.8 | 1.25 | 0.7 | 0.161 | 0.50 | 5.86 | 0.79 | 7.4 |
| 19 | 0.823 | 7.8 | 70.9 | 2.58 | 1.0 | 0.177 | 0.62 | 5.81 | 0.93 | 6.2 |
| 20 | 1.100 | 3.5 | 5.9 | 0.26 | 0.5 | 0.161 | 0.28 | 4.82 | 1.04 | 4.6 |
| 21 | 1.261 | 4.7 | 6.0 | 0.20 | 0.5 | 0.139 | 0.16 | 4.36 | 1.19 | 3.7 |

The mean value of the similarity criterium $\Pi_1$ is 0.163±0.009 as it is present at Fig.1 depending on its relative age (see below). The error is at 67 per cent of the probability with the minimal value of 0.139 for the most distant object and the maximum value of 0.183 for the cluster №9, the second in its luminosity at the Table 1, but small size and mass. For the monoatomic gas with the adiabate exponent 5/3 one can obtain that the sound velocity $c_e$ =0.75 $U_T$. Then the Mach number will be Ma ≈ 0.217±0.012 ≈ 0.22±0.01. This value is difficult to calibrate with any real measurements data, because in the real cluster of a complicated configuration and in disk-like systems, as our Galaxy, the velocity distributions are somewhat dependent on the distance from the gravity center. However the order of magnitude in hundreds of kilometers, as it present by eq. (5'), is quite close to what we know on the motion velocity in such systems. Let us say it again that the similarity parameter $\Pi_1$ and its near constancy with a scatter of less than 6 per cent, as at Fig.1 is in accord with the Tully–Fisher law. More precisely, it gives to the law a dimensional explanation: $L = G^{-1}U^5$, converting it to the physical relationship waiting for a more strict theoretical derivation. The numerical models, mentioned above, have not revealed it despite many details and complexities in them.

Fig.2 illustrates further the small variability of the similarity criterium $\Pi_1$. It shows a good proportionality of our two velocity scales in linear coordinates with the determination coefficient $R^2$ =0.91.

Table 2. Parameters of nearby clusters and $\Pi_2$.

| № | $z$ | $R_{500}$, Mpc | $T$ , keV | $M_{500}$ $10^{14}M_\odot$ | $\Pi_2$ | $T_a$, By | $T_d$, By | $T_a/T_d$ |
|---|---|---|---|---|---|---|---|---|
| 22 | 0.0569 | 1.007 | 4.14 | 3.17 | 2.58 | 11.3 | 2.58 | 4.38 |
| 23 | 0.0162 | 0.650 | 2.08 | 1.0 | 2.51 | 11.3 | 2.53 | 4.46 |
| 24 | 0.1883 | 0.944 | 4.81 | 3.06 | 2.28 | 11.6 | 2.64 | 4.4 |
| 25 | 0.0881 | 1.337 | 7.94 | 7.68 | 2.45 | 11.5 | 2.74 | 4.19 |
| 26 | 0.1603 | 1.096 | 5.96 | 4.56 | 2.37 | 11.6 | 2.67 | 4.35 |
| 27 | 0.1429 | 1.299 | 7.38 | 7.57 | 2.68 | 11.6 | 2.67 | 4.34 |
| 28 | 0.0622 | 1.235 | 6.12 | 6.03 | 2.70 | 11.4 | 2.73 | 4.18 |
| 29 | 0.0592 | 1.362 | 8.47 | 8.01 | 2.35 | 11.3 | 2.72 | 4.16 |
| 30 | 0.2302 | 1.416 | 8.89 | 10.74 | 2.89 | 11.7 | 2.57 | 4.55 |
| 31 | 0.0199 | 0.634 | 1.64 | 0.77 | 2.51 | 11.3 | 2.62 | 4.31 |

The Table 2 is a combination of Tables 4 and 9 from Vikhlinin et al (2006). From the Table 9 we exclude clusters with undetermined parameters necessary for testing our formulas.

The second similarity parameter $\Pi_2$, eq. (8), we write down as

$$\Pi_2 = 3.4 \ M/R \ T_e \qquad (9)$$

With the use of the measurement units above, the universal constant $G = 6.672 \cdot 10^{-12} \ \text{m}^3\text{kg}^{-1}\text{s}^{-2}$ and the value of the first similarity parameter $\Pi_1 = U_d/U_T = 0.163$, because of the absence directly measured velocities. Obtained in such a way values of the virial criteria $\Pi_2$ are introduced into the Tables 1 and 2. The numerical coefficient in (9) here contain $G$ and $\Pi_1$.

The Table 2 is an abbridged continuation of the Table 1 for nearby clusters with $z \leq 0.23$. Vikhlinin et al (2006) do not present luminosities, therefore there is no values of the first similarity criterium $\Pi_1$. To estimate the virial similarity parameter $\Pi_2$ we use the same value of $\Pi_1 = 0.163$ as for the Table 1. The values of the virial parameter $\Pi_2$ in this Table 2 are scattered much less than in the Table 1. The range for it is $2.18 \leq \Pi_2 \leq 2.89$ with the mean value $2.50 \pm 0.17$ or about 7 per cent, while for $z \geq 0.4$ the maximal value of $\Pi_2$ is 2.34 and the minimal value is 0.29. In both Tables we present ages, $T_a$, for the clusters in billion years, By. The age is estimated by dividing the age of the universe, 13.7 By, by $(1+z)$ and extracting from this result 1.7 By after the Big Bang when galaxies started to form (Rees, 2002). The dynamic time scale $T_d$, eq. (1ʹ), is used to normalized the cluster age.

There is question, what is the age to reach a virial equilibrium and at what value of the similarity parameters $\Pi_2$. For a system of material particles in the centrally–symmetrical gravity field the doubled kinetic energy is equal to potential energy, or to the depth of the potential well. Then $\Pi_2 = 1$, more precisely to -1 (Landau and Lifshits, 1960). The gravity field of a cluster consisting of a multitude of galaxies has a complicated spatial structure, therefore one may expect that then $\Pi_2 \geq 1$. Therefore we shall assume that ten nearby clusters at the Table 2 are all virialized. For them the criterium $\Pi_2$ are in a narrow range around $\Pi_2 \approx 2.5$. This formally means that the potential well is so deep that the potential energy is fivefold larger than the kinetic one.

Fig.2 presents the values of the virial criterium $\Pi_2$ for 30 clusters in dependence on their relative age $T_a/T_d = \tau$. This shows how many cycles of mixing the given cluster had survived before its radiation could reach the Earth. For the nearby clusters of the Table 2 the relative time $\tau$ is of the order 4 with a mean value $\tau = 4.13 \pm 0.12$. For faraway cluster of the Table 1 at $z \geq 0.4$ the value of $\Pi_2$ is changing from 2.34 until 0.29 and $\Pi_2 \leq 1$ for clusters at $z > 0.6$ and for them the relative time $\tau \leq 3$. Therefore for the virial relaxation of the clusters the relative time $\tau$ has to be not less than 3 or even larger. This situation reminds us the collisionless plasma where the relaxation time is of several characteristic periods (Kadomtsev, 1976) while in the gas kinetic theory the thermodynamic equilibrium is established during one or two molecular collision times.

An attentive look at cluster masses in our Tables reveals that with rare exceptions the masses of distant clusters are significantly smaller than for nearby ones and the masses on average increase with cluster age $T_a$, as well as the dynamic times $T_d$. This means that clusters already relaxed continue to absorb new galaxies and even smaller clusters. This increases their virial coefficient $\Pi_2$ and the dynamic time $T_d$, see further below the end of the section 3.

### 3. Tests oh the data quality and relationship between the measured values

The situation with our knowledge of the cluster parameters is in a sense unique in natural sciences. Nowhere we know so many measurable and measured characteristics as here: mass, luminosity, size, velocity, and temperature united by gravity field. This gives possibilities to construct by dimension considerations many relationships among various characteristics and to test them at once with estimates of the numerical coefficients in those relationships. These numerical coefficients may serve as modified similarity criteria, of course, being functions of the two above $\Pi_1$ and $\Pi_2$.

We start with tests of the data quality and mutual adjustment. The initial test is in the estimate of the universal gravity constant $G = 6.672 \cdot 10^{-12} \text{ m}^3\text{kg}^{-1}\text{s}^{-2}$. We use first the Tully–Fisher law in our interpretation:

$$G = c_g U_d^5 L^{-1} \qquad (10)$$

with $c_g$ as a numerical coefficient to be determined. The values of luminosity $L$ is only for 21 object in Table 1. However there are no velocities there but temperature instead. Therefore we may use the thermal velocity (2) and the first practically constant similarity criterium $\Pi_1 = U_d/U_T = 0.163$. Taking into account eq. (2) we may transform eq. (10) using our measurement units into:

$$10^{12}G = 1.84\, c_g T_e^{5/2} L^{-1} . \qquad (11)$$

Using first eleven objects of the Table 2 the r.h.s. of the equation is found to be 6.88 ±1.94. Therefore the dimensional adjustment coefficient $c_g = G/G_{exp} = 0.97$. The scatter of about 28 per cent is caused first of all by a high value of the exponent at the temperature.

The second method is based on another dimensional formula $G = (R/M)^{5/3}L^{2/3.}$ (see below eq.(13)), which in our measurement units can be written as

$$10^{12}G = 2.06\, c_g (R/M)^{5/3} L^{2/3} \qquad (12)$$

with coefficient 2.06 arising from conversion of our special measurement units to SI, System international. The r.h.s. gives 2.5 $c_g$ (1±0.2) and $c_g = G/G_{exp} = 2.67$ for the first eleven objects of the Table 1 (2.5 $c_g = 6.675$). Therefore with the scatter 20-28 per cent we are able to estimate the universal constant $G$. The scatter gives a feeling on the data quality: it is within 30 per cent or better.

We present several more relationship among the measured variables, some of which have been already described by the aforementioned reviews. First of all the relationship between luminosity and temperature which is different from the Stefan-Boltzmann $\sigma T_e^4$ due to seemingly small gas density of thermodynamic equilibrium between plasma and its X-ray radiation. This follows from Tully – Fisher law in the form of eq.(11). Our Fig.3 is taken from Voit (2005), where it is Fig. 14 to which we add the line with the 5/2 slope at the temperatures $T > 1$ keV.

At smaller temperatures the slope is steeper, close to $T^4$, which may be an evidence for a denser plasma at $T_e \leq 10^7$ K with a local thermodynamic equilibrium between plasma and radiation. The data scatter there is extremely large and some new effects may appear there together with additional similarity criteria.

Quite popular at the studies is the mass-temperature dependence which is usually approximated as a power law: $M \approx T^n$. If the plasma is in a hydrostatic equilibrium then in a centrally – symmetric gravity field $n = 1.5$ (see Voit, 2005). Vikhlinin et al (2006) on the base of nearby clusters, $z \leq 0.23$, propose $n = (1.5 \div 1.6) \pm 0.1$ and at Saunderson et al (2003) $n = 1.80 \pm 0.06$. A closer analysis of the Fig.15 from Voit (2005) shows that the exponent $n = 1.5$ describes well the data at $T \geq 2$ keV, and the larger values of $n$ appear at the inclusion of the temperatures $T_e \leq 2$keV. Voit (2005) at his Fig.5 presents a linear dependence $L \sim M\,T_e$. With the just discussed dependence $M \sim T^{3/2}$ it gives $L \sim T^{5/2}$ as in our dimensional formula, a consequence of the Tully–Fisher law, presented by the direct line at our Fig.3.

Because we have data on both luminosity and temperature for clusters in the Table 1 we could establish directly a power law correlation between these two variables considered as stochastic ones. The regression line has a mean slope $n = 2.37 \pm 0.39$ with 95 per cent of probability. The scatter is increased by a narrow range of temperatures, less than one order of magnitude (2.2 to 14.1 keV) and a limited number of points. Nevertheless it envelops well the dimensional exponent $n = 2.5$.

Note also the dependence of luminosity from mass and size. The dimensional considerations produce:

$$L = c_l\,(M/R)^{5/2}\,G^{3/2}, \tag{13}$$

where $c_l$ is a numerical coefficient. Because here are global quantities: mass and size, it should be applied to already virialized, or relaxed, objects. Therefore for this purpose we have only first eleven objects from Table 1, which, one may hope, are already relaxed. From eq. (13) in our units of dimension $c_l = 0.1716\,G^{-3/2}L\,(R/M)^{5/2}$. First eleven objects of the Table 1, excluding object $N_0$ 9, give $c_l = 0.215 \pm 0.041$ with 67 per cent of probability. The scatter of about 20 per cent is caused by a high value of the exponent 5/2 at relatively poorly known size $R$. The object №9 gives $c_l = 0.606$ exhibiting that its luminosity is about three times fainter for its mass and size as it gives eq.(13). The other ten far objects of the Table 1 gives a large scatter for $c_l$ from 0.35 to 10 showing that they do not produce supposedly full possible luminosity as the already relaxed clusters. This reflects the gradual increase of the virial similarity parameter $\Pi_2$ with increase of the cluster age, see Table 1.

The other test for the luminosity follows from the Tully-Fisher relationship transformed into dependence on the temperature $T_e$ with eq. (2) using the thermal velocity instead of dynamical one of eq.(5). It gives:

$$L = c_1\,\Pi_1^5\,T_e^{5/2}G^{-1} \tag{14}$$

where $\Pi_1$ is the first similarity criterium, the ratio of dynamical velocity scale to the thermal one and $c_1$ is the numerical coefficient. In our dimension units:

$$L = c_{11} \; 0.275 \; T_e^{5/2} \tag{15}$$

There is a scatter in the values of the numerical coefficient determined from the Table 1 with $0.46 < c_{11} < 1.78$ with the mean value of $1.04 \pm 0.28$. If we exclude from consideration the outlier object № 9, then $c_{11} = 1.00 \pm 0.17$. The close proximity of the numerical coefficient $c_{11}$ to unity is an evidence of the correct value of the first similarity criterium $\Pi_1$ among all our data of measurements.

Dependence of the cluster mass from its temperature follows from the hydrostatic approximation by Kaiser (1986) the isothermal approximation: $M \sim T^{1.5}$ (see Vikhlinin et al , 2006; Saunderson et al , 2003). In these two papers it is proposed that $n = 1.5 - 1.85$. From the dimensional arguments it follows in our units of dimension:

$$M = c_m R \; T_e \tag{16}$$

Because the size $R$ also depends on the temperature the total exponent $n$ could also be larger than 3/2. The difference is a reflection of the continuing virialization of the clusters from $z \sim 0.5$ towards $z \leq 0.23$ both by the increasing mass and , to a smaller extent, of size. This process is revealing itself by increase of the similarity criterium $\Pi_2$. Fig.5 presents a scatter plot of observed masses vs their dimensional expressions as in eq. (16), triangles are for the relaxed objects of the Table 2 and full circles are for the first 11 objects from the Table 1. Note high values of the determination coefficient $R^2$ in the linear scales, not logarithmic ones.

The size of the clusters is also increasing with their age but clearly , to a smaller extent than their mass. This process is reflected in the evolution of the ratio $M/R$ at the virial criterium $\Pi_2$. This ratio depends only on the temperature, see eq. (8) and (3), as

$$M/R = 0.295 \; r_2 \; T_e \tag{17}$$

in our units and $r_2$ is a numerical coefficient. For the first 11 objects of the Table 1 we have $r_2 = 1.83 \pm 0.23$ at $0.4 \leq z < 0.6$ and $r_2 = 2.50 \pm 0.17$ for the nearby relaxed objects of the Table 2.

It is worthwhile to propose a dimensional estimate of the total energy of a cluster using its other measured characteristics:

$$E = M^2 G^{-1} R^{-1} \tag{18}$$

which in our system of units is

$$E = 8.7 \cdot 10^{54} M^2 R^{-1} \approx 10^{55} M^2 R^{-1}, \text{Jouls} . \tag{18'}$$

From the other hand the total heat content, or its enthalpy is due to eq.(4') equal to $F \approx 1 \cdot 10^{35} M \; T_e$, Jouls. For the thirty objects of our two Tables this heat energy changes from 1.87 to 87 in the units $10^{55}$ J. For the first eleven supposedly virialized objects of the Table 1 the enthalpy in the same units changes from 4.8 to 150. For a comparison the total released energy for the supernova explosions is in the range $10^{44} - 10^{45}$ J, or about eleven-twelve order of magnitude smaller.

We may consider here the related quantities – the pressure and particle concentration in clusters. We start with short consideration of the thermodynamic equilibrium in cluster gas, just for the convenience of readers (detailed considerations are in Kaiser, 1985). From mass and mean molecular weight of $\mu = 0.6$ we find to total particle concentration in a cluster as $N_0 = M/\mu m_p$ and their volume concentration as

$$N = N_0 R^{-3} = M/\mu m_p R^3 = 0.68 \cdot 10^{-12} M R^{-3} \tag{19}$$

in our system units adjusted to SI. For 10 relaxed clusters of the Table 2 we have $10^{-4} N = 2.26 \pm 0.13$ at min 2.11 and max 2.47. Eleven first objects of the Table 1 have $10^{-4} N = 5.12 \pm 1.66$ at min 2.50 and max 8.19. The much larger scatter and greater values of the concentration number $N$ is, evidently, caused by different procedures of the size determinations for distant and nearby clusters. Anyway it would be safe to accept that $N \sim 10^4 \mathrm{m}^{-3}$. With the scattering cross-section of $10^{-20}$ m$^2$ (or 1 sq. angstrom, for plasma it could be larger) we have the particle free path $\lambda = (N\sigma)^{-1} \sim 10^{16}$ m, or a third of the parsec, or one light year. Therefore at 1Mpc it would be 3 million interactions – scattering which is a large excess of time for establishing the thermodynamic equilibrium. But it looks like it is not enough for an equilibrium between plasma and X-ray radiation, in other words , plasma could be optically thin for the radiation and we have luminosity $L \sim T^{5/2}$ at $1 < T_e < 10$ keV instead of $T^4$ as it should be for the black body emissivity due to the Stefan-Boltzmann law. At smaller temperatures there is much larger scatter of the data at Fig.2 with decreasing luminosity. Just for comparison we put $L \sim T^4$ as the dashed line.

Another quantity interesting in this respect is the pressure considered as the volume energy density. From dimensional arguments, compare with eq. (18), we have

$$p_d = E/V = (M^2 \cdot G)/R^4 \tag{20}$$

which should be compared to the kinetic theory expression $p_t = NkT$. Comparing this two expressions we have a new similarity criterium according for (3)

$$\Pi_2' = p_d/p_T = [MG/RT_0] \cdot [\mu m_p/kA] = 0.27 \, M/(R \, T_e) \tag{21}$$

where $\mu m_p = 10^{-27}$ kg, $k = 1.38 \cdot 10^{-23}$ J/K, A $= 1.161 \cdot 10^7$ K/eV and 0.27 appears after conversing our system of units to SI. The new criterium is related to the virial criterium as $\Pi_2' = 3 \, \Pi_1^2 \, \Pi_2$. With $\Pi_1 = 0.163$ as above we obtain:

$$\Pi_2 \approx 12.5 \, \Pi_2' \tag{22}$$

The values of this criterium $\Pi_2'$ vary from 0.174 to 0.230 with the mean $0.199 \pm 0.014$ with 67 per cent probability for the nearby clusters at $z \leq 0.2$ in the Table 2. For first eleven clusters, supposedly relaxed, of the Table 1 it is $\Pi_2' = 0.146 \pm 0.018$ showing again that these clusters are undersuperrelaxated in the virial sense as the nearby clusters.

Other dimensional formulas are acting not badly but the numerical coefficients are somewhat depending on the agree of virialization $\Pi_2$. We illustrate this by a couple of examples. First one is for the mass $M_{dim} = L^{2/5} R G^{-3/5}$ which assuming the preservation of the first criterium $\Pi_1$ may be transformed to the eq. (16) and the results are presented at Fig.6. We check it simply by introducing the numerical coefficient $c_m = M_{obs} / M_{dim}$ and calculate it separately for first 11 objects of the Table 1 and for the 10 objects of the Table 2. In the first case $c_m = 0.182 \pm 0.19$, and in the second case $c_m = $ 1.32 ± 0.11 which corresponds quite well to the results at Fig. 6 demonstrating the fact that the similarity criterium $\Pi_1$ is almost permanent indeed.

The dimensional expression for the cluster size is

$$R_{\dim} = MG^{3/5}L^{-2/5} \qquad (23)$$

which with eq.(14) can be transformed into $R = c_r M / T_e$ at the same assumption of the near constancy of $\Pi_1 = 0.163$. The test is for the relaxed nearby clusters of the Table 1 which are determined with a better precision. The scatter plot is at Fig.6.

A peculiar form has the equation of state for gas of clusters considered as an ideal one. We start with the total energy of the cluster presented by eq.(18). The energy per unit volume is the pressure

$$p = M^2G/ R^4 = \rho^2 GR^2 \qquad (24)$$

The presence of the clusters size in thermodynamics here is a consequence of viriality. At the other hand for the ideal gas $p = nkT$ and density $\rho = \mu m_p = 1 \cdot 10^{-27}$ kg and $n$ is the volume concentration of the particle number. From these three formulas we find

$$\rho = k / \mu m_p G \cdot T_e / R^2 \qquad (25)$$

In our system of units: $T_e$ in keV, $R$ in $M$pc we find that

$$\rho = 2.5 \cdot 10^{-23} T_e R^{-2}, \text{kg/m}^3 \qquad (26)$$

A numerical coefficient may be present here, hopefully of order unity, but there are no data to determine it. If so, the number density for our clusters can be estimated as $10^4 - 10^5$ of particles per cubic meters. The total number of particles could be of order $10^{72}$ with uncertainty of couple of order magnitude. One should note that for virialized objects $R \sim T^{-1}$ and therefore there is nothing strange in (25), which means that colder gas is a denser one.

## 4. Discussion of the results

We want to repeat it again that in the case of the clusters we have a unique situation in science: a redundancy of measured characteristics which allows one to check an internal consistency of the measured quantities and perform a substantial analysis of this situation. The four measured characteristics: luminosity, mass, size, and velocity, or temperature, binded by gravitation , i.e. by the universal gravity constant $G$, are all characterized by only three dimension units : mass $M$, length $L$, and time $T$. Therefore, by the celebrated $\Pi$-theorem one can form from these five quantities two non-dimensional parameters, usually called the similarity criteria. Also dimensional scales of the quantities can be formed .With temperature the thermal velocity of gas particles can be formed (Rees, 2002). The luminosity and gravity constant produce the dynamic velocity scale $U_d = (GL)^{1/5}$ , eq. (4) and (4') which for clusters and separate galaxies is of order hundreds kilometers per second. Its ratio to the thermal velocity generates the first similarity criterium $\Pi_1$. Its value is found nearly constant for the 21 cluster of Table 1: $\Pi_1 = 0.163 \pm 0.009$ with 67 per cent probability. The small variability of this similarity criterium explains at least phenomenologically, the celebrated Tully – Fisher (1977) empirical law, initially found in radioastronomy.

The second criterium $\Pi_2 = MG/U^2R$, eq.(8), is the virial relationship, the ratio of potential energy to the doubled kinetic energy of the system. For the centrally-symmetric configuration $\Pi_2 = 1$ at the final stage of the system evolution. There is the dynamic time scale (Rees, 2002) $T_d$, eq. (1). The estimates of the criterium $\Pi_2$ for 31 clusters of the Table 1 and 2 reveal that $\Pi_2$ is increasing with the cluster age from about 0.3 at $z = 1.26$ to 2.5 at $z \leq 0.23$. This is the direct evidence of the evolution of the galactic clusters observed with our own eyes due to fivefold increase of masses and threefold increase of the ratio $M/R$ entering the virial criterium $\Pi_2$. Two our Tables and Fig.1 show that to reach some virial equilibrium at least three or four dynamic times $T_d$ are necessary for clusters which amounts to several billion years . The evolution is slowing but there is no clear sign that it is finishing after, say 12 By of the age of the nearby clusters.  To clarify this it is necessary to perform a statistical analysis of cluster mass distributions at, say, $z = 0.05, 0.1, 0.2$ etc, which should reveal the details of the cluster virial  evolution, their mass distribution and its homogeneity in time and space.

The nearby constancy of the first similarity criterium $\Pi_1$ is a formalization of the empirical Tully – Fisher law through the dynamical velocity scale $U = (LG)^{1/5}$, eq. (5), and the particle thermal velocity, eq. (2). Understood now dimensionally the Tully – Fisher law should be also obtained "from the first principles" which could be a long way, even numerically, see Voit (2005), Katsov and Borrega (2012).

The second similarity criterium $\Pi_2$ shows a degree of the virial equilibrium between the potential energy, or a depth of the gravity well, and kinetic energy of the motions. Distant clusters at $z \geq 0.7$ demonstrate clearly their underrelaxation in this sense. Closer clusters show an increase of a potential well, or the criterium $\Pi_2$. For nearby clusters the potential energy is about fivefold of the kinetic one while for the remote ones the situation it reversed, see Table 1. This is a direct evidence of the evolution of our Universe, revealing how clusters increase their mass at the expense of minor ones – "cannibalism"! The nearby clusters are on average several times more massive than remote ones.

It is unfortunate that total luminosities are presented, together with other characteristics, only by Vikhlinin  (2002), or, at least, this author wasn't able to find them.  Therefore I could not estimate directly the similarity criterium $\Pi_1$ ,  except for 21 cluster of the Table 1. But the fact that $\Pi_1$ for the Sun is only twice smaller than for the nearby and distant clusters suggests that it is an important characteristics with supposedly small scatter for many classes of the objects. And this should be explored widely.  I hope the younger people will do this.

**Acknowledgements**

I am greatly indebted to Rashid Syunyaev for convincing me to be a reviewer of Dr. Sci. Dissertation of  A.Vikhlinin  on galactic clusters in 2001, for Alexey Vikhlinin answering  my various questions and supplying me with additional materials, for Grigorii Falkovich who invited me in 1999 to the Weizmann Institute for a couple of weeks where I had heard for the first time about Tully – Fisher law, was struck by its outragions  simplicity and saw at once that it needs $G$. Many other administrative and scientific duties distructed me from the subject, primarily I am geophysical fluid dynamicist, until December 31, 2008. Then I finished to be my Insitute director, and I seriously started to complete the book (Golitsyn, 2012b). One paragraph of it is on galaxies and clusters. Out of it I wrote a paper (Golitsyn, 2012a) with part of results presented here. At the end of October 2012 in Vienna I had a short encounter with Martin Rees who asked me where he could read what I had informed him briefly on the subject. So this text is a fulfillment of what I had promised to him. Recent discussions with Olga Silchenko and Rashid Syunyaev were also stimulating and useful.  Dr. Otto Chkhetiani, an astronomer by education, was a great help for me during the work on the subject.

## Figure captions

Fig.1. The similarity criterium  $\Pi_1 = U_d/U_T$     for 21 clusters from the Table 1 in dependence on the cluster age in terms of dynamical time scales  $T_d$ as in eq. (1).

Fig.2. The linear proportionality of our two velocity scales.

Fig.3. Virial   similarity criterium  $\Pi_2 = MGU^{-2}R^{-1}$,  eq.(8) and (9),  for 31 clusters from the Tables 1 and 2  in dependence on the cluster age as at Fig.1.

Fig.4.Clusters luminosities of real ones – large symbols, and model ones – small dots from Voit (2005) where it is Fig.14. Thin line corresponds to eq. (13) -  $L \sim T^{5/2}$, and dashed line is $L \sim T^4$ .

Fig.5. Scatter plot of  the  luminosity dependence on the temperature according to the Table 1. The regression line in logarithmic coordinates has a slope $n = 2.37 \pm 0.39$ with 95 per cent of probability, $R^2 = 0.724$.

Fig.6. Scatter plot for masses from the Table 1 against their dimensional expressions as in eq. (16), well relaxed objects from the Table 2, • - first eleven objects from the Table 1,  eq.  (24); line1 -  for  $z \geq 0.4$, line 2 – for $z \leq 0.23$.

Fig.7.  Scatter  plot  of the cluster size from the Table 2  against their dimensional expression by eq. (23).

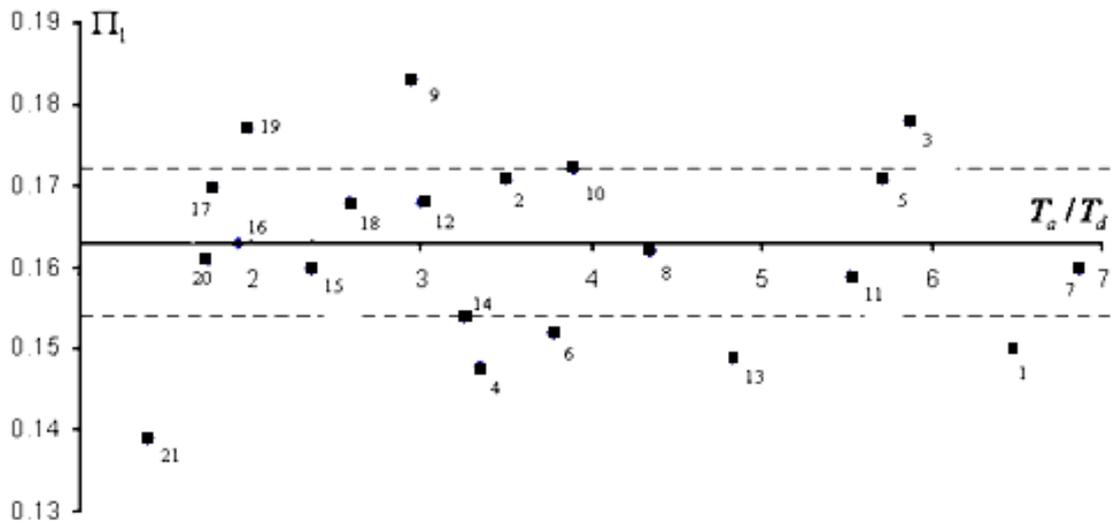

Fig. 1 The similarity criterium $\Pi_1 = U_d/U_T$ for 21 clusters from the Table 1 in dependence on the cluster age in terms of dynamical time scales $T_d$ as in eq. (1).

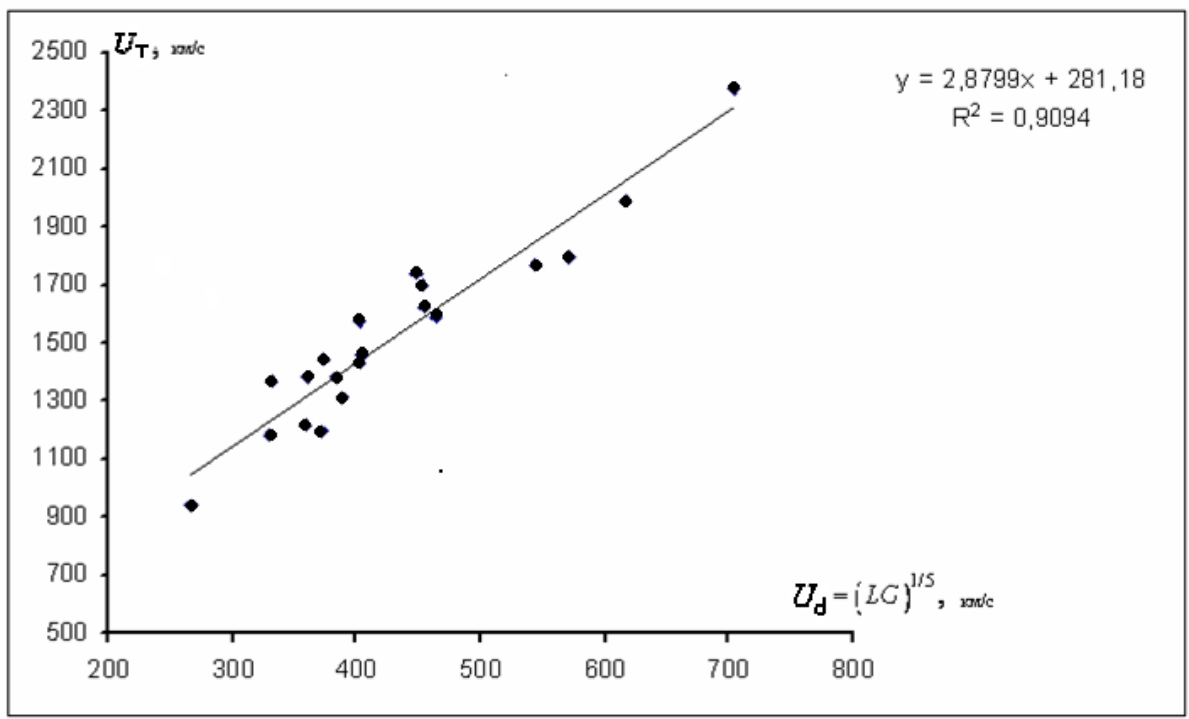

Fig. 2 The linear proportionality of our two velocity scales.

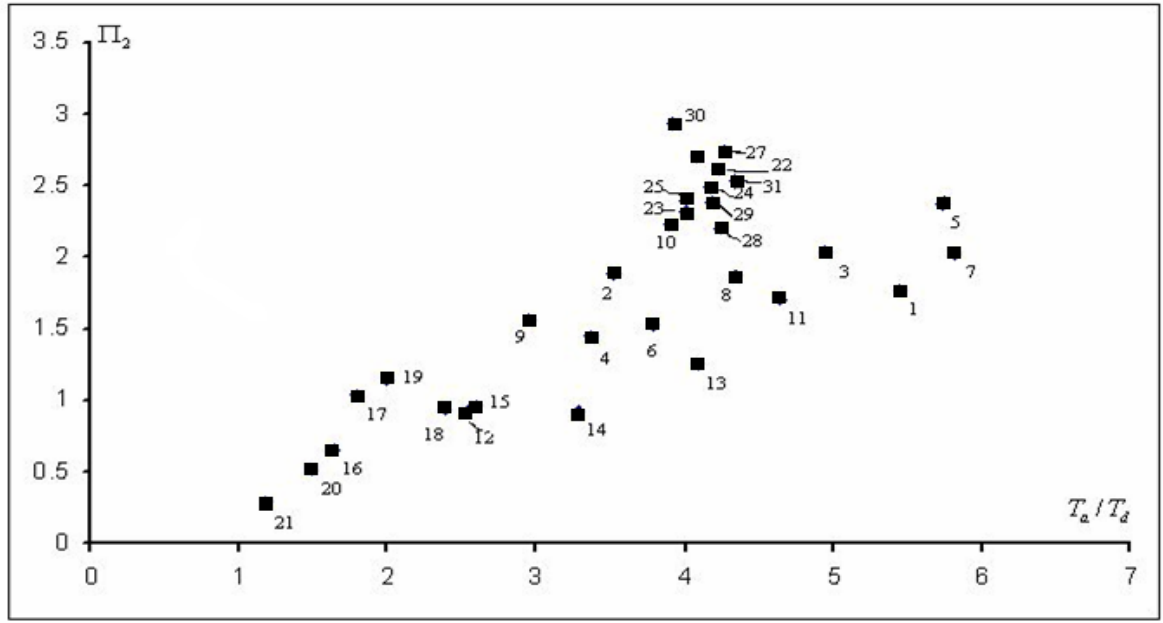

Fig. 3 Virial similarity criterium $\Pi_2 = MGU^{-2}R^{-1}$, eq.(8) and (9), for 31 clusters from the Tables 1 and 2 in dependence on the cluster age as at Fig.1.

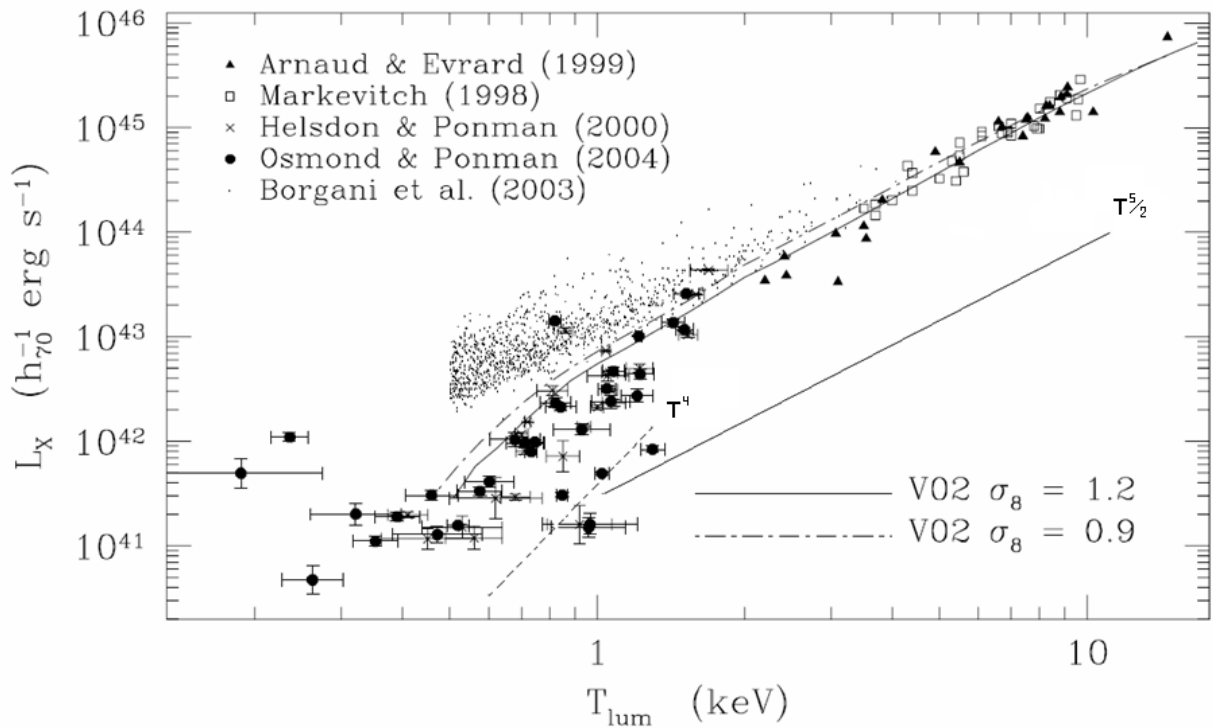

Fig. 4 Clusters luminosities of real ones – large symbols, and model ones – small dots from Voit (2005) where it is Fig.14. Thin line corresponds to eq. (13) - $L \sim T^{5/2}$, and dashed line is $L \sim T^4$.

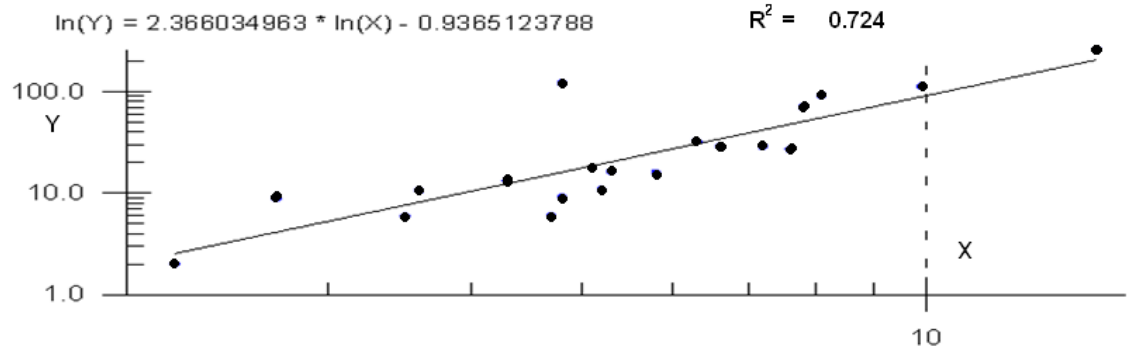

Fig. 5 Scatter plot of the luminosity dependence on the temperature according to the Table 1. The regression line in logarithmic coordinates has a slope $n = 2.37 \pm 0.39$ with 95 per cent of probability, $R^2 = 0.724$.

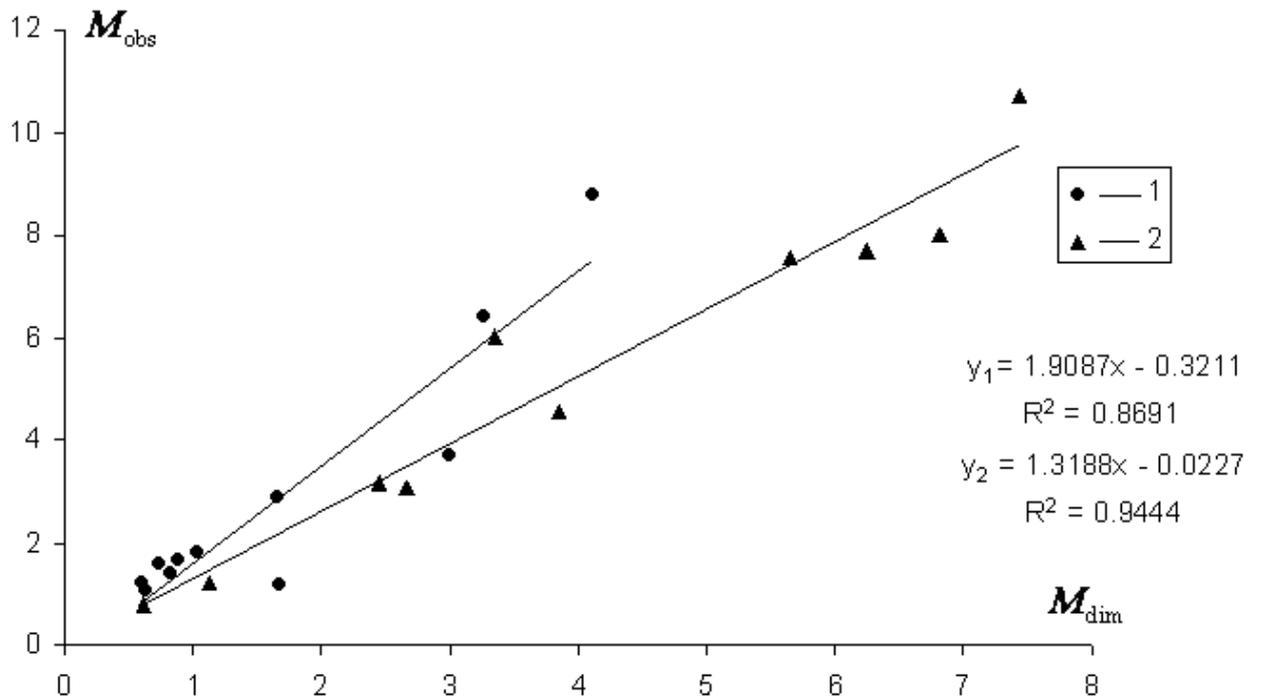

Fig. 6 Scatter plot for masses from the Table 1 against their dimensional expressions as in eq. (16), well relaxed objects from the Table 2, • - first eleven objects from the Table 1, eq. (24); line1 - for $z \geq 0.4$, line 2 – for $z \leq 0.23$.

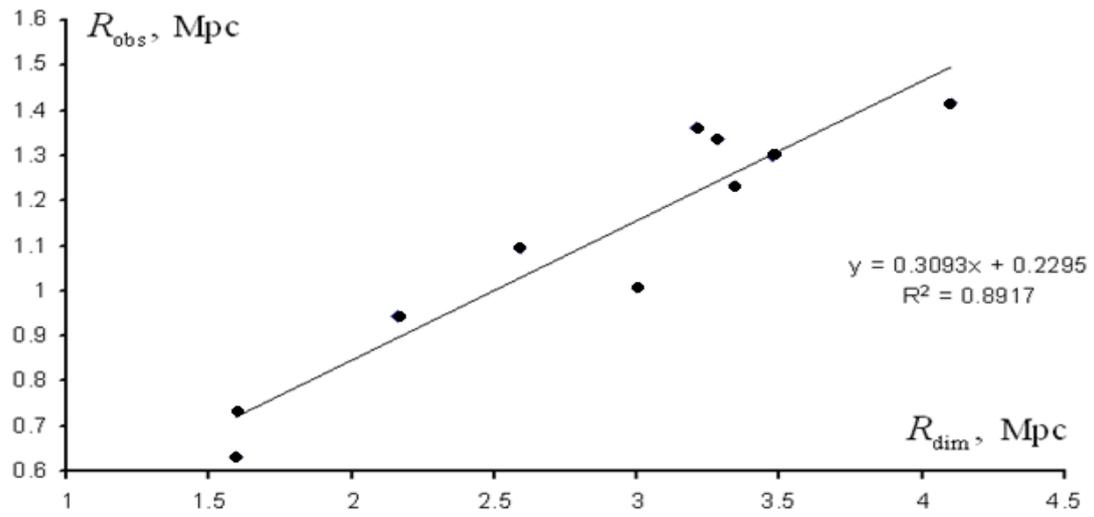

Fig. 7 Scatter plot of the cluster size from the Table 2 against their dimensional expression by eq. (23).